\documentclass[12pt]{article}
\newcommand{\nc}{\newcommand}
\nc{\renc}{\renewcommand}

\usepackage{calc}
\usepackage{ifthen}
\usepackage{graphicx}
\usepackage{epsfig}
\usepackage{miniplot}

\usepackage{wrapfig}
\usepackage{subfigure}
\usepackage{rotfloat}

\usepackage{epsfig}
\usepackage{graphicx}
\usepackage{epsf}

\nc{\half}{{\textstyle{1\over2}}}
\nc{\etal}{\mbox{\it et al. }}
\nc{\ie}{{\it i.e.}}
\nc{\eg}{{\it e.g.}}

\renc{\thefootnote}{\arabic{footnote}}
\nc{\capt}[1]{{\bf Figure.} {\small\sl #1}}

\nc{\eqs}[2]{\mbox{Eqs.~(\ref{#1},\,\ref{#2})}}
\nc{\eq}[1]{\mbox{Eq.~(\ref{#1})}}

\nc{\figs}[2]{\mbox{Figs.~(\ref{#1},\,\ref{#2})}}
\nc{\fig}[1]{\mbox{Fig~.(\ref{#1})}}

\nc{\tag}[1]{\label{#1} \marginpar{{\footnotesize #1}}}
\nc{\mtag}[1]{\label{#1} \mbox{\marginpar{{\footnotesize #1}}}}
\renc{\baselinestretch}{1.5}
\newlength{\overeqskip}
\newlength{\undereqskip}
\nc{\be}[1]{\begin{equation} \mbox{$\label{#1}$}}
\nc{\bea}[1]{\begin{eqnarray} \mbox{$\label{#1}$}}
\nc{\Section}[2]{\section{#2}\label{#1}}
\nc{\Bibitem}[1]{\bibitem{#1}}
\nc{\Label}[1]{\label{#1}}

\nc{\eea}{\vspace{\undereqskip}\end{eqnarray}}
\nc{\ee}{\vspace{\undereqskip}\end{equation}}
\nc{\bdm}{\begin{displaymath}}
\nc{\edm}{\end{displaymath}}
\nc{\dpsty}{\displaystyle}
\nc{\bc}{\begin{center}}
\nc{\ec}{\end{center}}
\nc{\ba}{\begin{array}}
\nc{\ea}{\end{array}}
\nc{\bab}{\begin{abstract}}
\nc{\eab}{\end{abstract}}
\nc{\btab}{\begin{tabular}}
\nc{\etab}{\end{tabular}}
\nc{\bit}{\begin{itemize}}
\nc{\eit}{\end{itemize}}
\nc{\ben}{\begin{enumerate}}
\nc{\een}{\end{enumerate}}
\nc{\bfig}{\begin{figure}}
\nc{\efig}{\end{figure}}
\nc{\arreq}{&\!=\!&}
\nc{\arrmi}{&\!-\!&}
\nc{\arrpl}{&\!+\!&}
\nc{\arrap}{&\!\!\!\approx\!\!\!&}
\nc{\non}{\nonumber\\*}
\nc{\align}{\!\!\!\!\!\!\!\!&&}

\def\lsim{\; \raise0.3ex\hbox{$<$\kern-0.75em
      \raise-1.1ex\hbox{$\sim$}}\; }
\def\gsim{\; \raise0.3ex\hbox{$>$\kern-0.75em
      \raise-1.1ex\hbox{$\sim$}}\; }
\nc{\DOT}{\hspace{-0.08in}{\bf .}\hspace{0.1in}}
\nc{\Laada}{\hbox {$\sqcap$ \kern -1em $\sqcup$}}
\nc\loota{{\scriptstyle\sqcap\kern-0.55em\hbox{$\scriptstyle\sqcup$}}}
\nc\Loota{{\sqcap\kern-0.65em\hbox{$\sqcup$}}}
\nc\laada{\Loota}
\nc{\qed}{\hskip 3em \hbox{\BOX} \vskip 2ex}

\nc{\real}{{\rm I \! R}}
\nc{\Z}{{\sf Z \!\!\! Z}}
\nc{\complex}{{\rm C\!\!\! {\sf I}\,\,}}
\def\bigid{\leavevmode\hbox{\small1\kern-3.8pt\normalsize1}}
\def\id{\leavevmode\hbox{\small1\kern-3.3pt\normalsize1}}
\nc{\slask}{\!\!\!/}
\nc{\bis}{{\prime\prime}}
\nc{\pa}{\partial}
\nc{\na}{\nabla}
\nc{\ra}{\rangle}
\nc{\la}{\langle}
\nc{\goto}{\rightarrow}
\nc{\swap}{\leftrightarrow}

\nc{\EE}[1]{ \mbox{$\cdot10^{#1}$} }
\nc{\abs}[1]{\left|#1\right|}
\nc{\at}[2]{\left.#1\right|_{#2}}
\nc{\norm}[1]{\|#1\|}
\nc{\abscut}[2]{\Abs{#1}_{\scriptscriptstyle#2}}
\nc{\vek}[1]{{\rm\bf #1}}
\nc{\integral}[2]{\int\limits_{#1}^{#2}}
\nc{\inv}[1]{\frac{1}{#1}}
\nc{\dd}[2]{{{\partial #1}\over{\partial #2}}}
\nc{\ddd}[2]{{{{\partial}^2 #1}\over{\partial {#2}^2}}}
\nc{\dddd}[3]{{{{\partial}^2 #1}\over
        {\partial #2 \partial #3}}}
\nc{\dder}[2]{{{d #1}\over{d #2}}}
\nc{\ddder}[2]{{{d^2 #1}\over{d {#2}^2}}}
\nc{\dddder}[3]{{d^2 #1}\over
        {d #2 d #3}}
\nc{\dx}[1]{d\,^{#1}x}
\nc{\dy}[1]{d\,^{#1}y}
\nc{\dz}[1]{d\,^{#1}z}
\nc{\dl}[1]{\frac{d\,^{#1}l}{(2\pi)^{#1}}}
\nc{\dk}[1]{\frac{d\,^{#1}k}{(2\pi)^{#1}}}
\nc{\dq}[1]{\frac{d\,^{#1}q}{(2\pi)^{#1}}}

\nc{\cc}{\mbox{$c.c.$ }}
\nc{\hc}{\mbox{$h.c.$ }}
\nc{\cf}{cf.\ }
\nc{\erfc}{{\rm erfc}}
\nc{\Tr}{{\rm Tr\,}}
\nc{\tr}{{\rm tr\,}}
\nc{\pol}{{\rm pol}}
\nc{\sign}{{\rm sign}}
\nc{\bfT}{{\bf T }}

\nc{\cA}{{\cal A}}
\nc{\cB}{{\cal B}}
\nc{\cD}{{\cal D}}
\nc{\cE}{{\cal E}}
\nc{\cG}{{\cal G}}
\nc{\cH}{{\cal H}}
\nc{\cL}{{\cal L}}
\nc{\cO}{{\cal O}}
\nc{\cT}{{\cal T}}
\nc{\cN}{{\cal N}}
\nc{\rvac}[1]{|{\cal O}#1\rangle}
\nc{\lvac}[1]{\langle{\cal O}#1|}
\nc{\rvacb}[1]{|{\cal O}_\beta #1\rangle}
\nc{\lvacb}[1]{\langle{\cal O}_\beta #1 |}
\nc{\bb}{\bar{\beta}}
\nc{\bt}{\tilde{\beta}}
\nc{\ctH}{\tilde{\cal H}}
\nc{\chH}{\hat{\cal H}}
\nc{\1}{\aa}
\nc{\2}{\"{a}}
\nc{\3}{\"{o}}
\nc{\4}{\AA}
\nc{\5}{\"{A}}
\nc{\6}{\"{O}}
\nc{\al}{\alpha}
\nc{\g}{\gamma}
\nc{\Del}{\Delta}
\nc{\e}{\epsilon}
\nc{\eps}{\epsilon}
\nc{\lam}{\lambda}
\nc{\om}{\omega}
\nc{\Om}{\Omega}
\nc{\ve}{\varepsilon}
\nc{\mn}{{\mu\nu}}
\nc{\vp}{\varphi}

\nc{\advp}[3]{{\it  Adv.\ in\ Phys.\ }{{\bf #1} {(#2)} {#3}}}
\nc{\annp}[3]{{\it  Ann.\ Phys.\ (N.Y.)\ }{{\bf #1} {(#2)} {#3}}}
\nc{\apl}[3]{{\it  Appl. Phys. Lett. }{{\bf #1} {(#2)} {#3}}}
\nc{\apj}[3]{{\it  Ap.\ J.\ }{{\bf #1} {(#2)} {#3}}}
\nc{\apjl}[3]{{\it  Ap.\ J.\ Lett.\ }{{\bf #1} {(#2)} {#3}}}
\nc{\app}[3]{{\it Astropart.\ Phys.\ }{{\bf #1} {(#2)} {#3}}}
\nc{\cmp}[3]{{\it  Comm.\ Math.\ Phys.\ }{{ \bf #1} {(#2)} {#3}}}
\nc{\cqg}[3]{{\it  Class.\ Quant.\ Grav.\ }{{\bf #1} {(#2)} {#3}}}
\nc{\epl}[3]{{\it  Europhys.\ Lett.\ }{{\bf #1} {(#2)} {#3}}}
\nc{\ijmp}[3]{{\it Int.\ J.\ Mod.\ Phys.\ }{{\bf #1} {(#2)} {#3}}}
\nc{\ijtp}[3]{{\it Int.\ J.\ Theor.\ Phys.\ }{{\bf #1} {(#2)} {#3}}}
\nc{\jmp}[3]{{\it  J.\ Math.\ Phys.\ }{{ \bf #1} {(#2)} {#3}}}
\nc{\jpa}[3]{{\it  J.\ Phys.\ A\ }{{\bf #1} {(#2)} {#3}}}
\nc{\jpc}[3]{{\it  J.\ Phys.\ C\ }{{\bf #1} {(#2)} {#3}}}
\nc{\jap}[3]{{\it J.\ Appl.\ Phys.\ }{{\bf #1} {(#2)} {#3}}}
\nc{\jpsj}[3]{{\it J.\ Phys.\ Soc.\ Japan\ }{{\bf #1} {(#2)} {#3}}}
\nc{\lmp}[3]{{\it Lett.\ Math.\ Phys.\ }{{\bf #1} {(#2)} {#3}}}
\nc{\mpl}[3]{{\it  Mod.\ Phys.\ Lett.\ }{{\bf #1} {(#2)} {#3}}}
\nc{\ncim}[3]{{\it  Nuov.\ Cim.\ }{{\bf #1} {(#2)} {#3}}}
\nc{\np}[3]{{\it  Nucl.\ Phys.\ }{{\bf #1} {(#2)} {#3}}}
\nc{\npps}[3]{{\it  Nucl.\ Phys.\ Proc.\ Suppl.\ }{{\bf #1} {(#2)} {#3}}}
\nc{\pr}[3]{{\it Phys.\ Rev.\ }{{\bf #1} {(#2)} {#3}}}
\nc{\pra}[3]{{\it  Phys.\ Rev.\ A\ }{{\bf #1} {(#2)} {#3}}}
\nc{\prb}[3]{{\it  Phys.\ Rev.\ B\ }{{{\bf #1} {(#2)} {#3}}}}
\nc{\prc}[3]{{\it  Phys.\ Rev.\ C\ }{{\bf #1} {(#2)} {#3}}}
\nc{\prd}[3]{{\it  Phys.\ Rev.\ D\ }{{\bf #1} {(#2)} {#3}}}
\nc{\prl}[3]{{\it Phys.\ Rev.\ Lett.\ }{{\bf #1} {(#2)} {#3}}}
\nc{\pl}[3]{{\it  Phys.\ Lett.\ }{{\bf #1} {(#2)} {#3}}}
\nc{\prep}[3]{{\it Phys.\ Rep.\ }{{\bf #1} {(#2)} {#3}}}
\nc{\prsl}[3]{{\it Proc.\ R.\ Soc.\ London\ }{{\bf #1} {(#2)} {#3}}}
\nc{\ptp}[3]{{\it  Prog.\ Theor.\ Phys.\ }{{\bf #1} {(#2)} {#3}}}
\nc{\ptps}[3]{{\it  Prog\ Theor.\ Phys.\ suppl.\ }{{\bf #1} {(#2)} {#3}}}
\nc{\physa}[3]{{\it  Physica\ A\ }{{\bf #1} {(#2)} {#3}}}
\nc{\physb}[3]{{\it  Physica\ B\ }{{\bf #1} {(#2)} {#3}}}
\nc{\phys}[3]{{\it Physica\ }{{\bf #1} {(#2)} {#3}}}
\nc{\rmp}[3]{{\it  Rev.\ Mod.\ Phys.\ }{{\bf #1} {(#2)} {#3}}}
\nc{\rpp}[3]{{\it Rep.\ Prog.\ Phys.\ }{{\bf #1} {(#2)} {#3}}}
\nc{\sjnp}[3]{{\it Sov.\ J.\ Nucl.\ Phys.\ }{{\bf #1} {(#2)} {#3}}}
\nc{\spjetp}[3]{{\it Sov.\ Phys.\ JETP\ }{{\bf #1} {(#2)} {#3}}}
\nc{\yf}[3]{{\it Yad.\ Fiz.\ }{{\bf #1} {(#2)} {#3}}}
\nc{\zetp}[3]{{\it Zh.\ Eksp.\ Teor.\ Fiz.\  }{{\bf #1}  {(#2)} {#3}}}
\nc{\zp}[3]{{\it Z.\ Phys.\ }{{\bf #1} {(#2)} {#3}}}
\nc{\ibid}[3]{{\sl ibid.\ }{{\bf #1} {#2} {#3}}}
\nc{\rf}[1]{(\ref{#1})}
\nc{\nn}{\nonumber \\*}
\nc{\bfB}{\bf{B}}
\nc{\bfv}{\bf{v}}
\nc{\bfx}{\bf{x}}
\nc{\bfy}{\bf{y}}
\nc{\vx}{\vec{x}}
\nc{\vy}{\vec{y}}
\nc{\oB}{\overline{B}}
\nc{\oI}{\overline{I}}
\nc{\oR}{\overline{R}}
\nc{\rar}{\rightarrow}
\nc{\ti}{\times}
\nc{\slsh}{\hskip-5pt/}
\nc{\sm}{Standard~Model~}
\nc{\MP}{M_{\rm Pl}}
\nc{\tp}{t_{\rm Pl}}
\nc{\ave}{\bar{E}}

\nc{\eff}{{\rm eff}}
\nc{\kk}{\vek{k}}
\nc{\pp}{{\rm p}}
\nc{\ga}{g_{a\gamma}}
\nc{\vv}{\\}
\nc{\eee}{{\bf E}}
\nc{\bbb}{{\bf B}}
\nc{\qcd}{T_{\rm QCD}}
\nc{\G}{\rm \ G}
\def\vec#1{{\bf #1}}

\def\lae{\;^{<}_{\sim} \;} \def\gae{\; ^{>}_{\sim} \;} 

\def\ell{e^{c}LL}

\begin{document}
{\title{\vskip-2truecm{\hfill {{\small \\
	\hfill \\
	}}\vskip 1truecm}
{\LARGE F-Term Inflation Q-Balls}}
{\author{
{\sc  John McDonald$^{1}$}\\
{\sl\small Cosmology and Astroparticle Physics Group, University of Lancaster,
Lancaster LA1 4YB, UK}
}
\maketitle
\begin{abstract}
\noindent

       A general analysis of Q-ball solutions
of the supersymmetric F-term hybrid inflation field equations is given. 
The solutions consist of 
a complex inflaton field and a real symmetry breaking field, with a conserved 
global charge associated with the inflaton. 
It is shown that the Q-ball solutions for any value of the superpotential coupling, $\kappa$, 
may be obtained from those with $\kappa = 1$ by rescaling the space coordinates.  
The complete range of Q-ball solutions for the case $\kappa = 1$ is given,
from which all possible F-term inflation Q-balls can be obtained. The possible 
role of F-term inflation Q-balls in cosmology is discussed.

\end{abstract} 
\vfil
 \footnoterule {\small $^1$j.mcdonald@lancaster.ac.uk}   
 \newpage 
\setcounter{page}{1}                   

\section{Introduction}

        Hybrid inflation models are widely regarded
as promising candidates for inflation in the context of 
supersymmetry (SUSY). There are two classes of SUSY hybrid inflation model \cite{lr}, 
F-term \cite{fti} and D-term inflation \cite{dti}, depending on which part of the scalar potential drives inflation. 
An important feature of all inflation scenarios is the process by which inflation ends and the Universe reheats. 
In SUSY hybrid inflation models this is likely to occur
via rapid tachyonic growth of sub-horizon quantum scalar field fluctuations into a space-dependent classical field, a process known as tachyonic preheating \cite{tp}. The subsequent evolution of the resulting inhomogeneous classical field is a complex numerical problem. In a recent simulation of the evolution of the scalar fields at the end of SUSY hybrid inflation \cite{mbs}, 
the formation of oscillons \cite{osc,osc2}  made of the inflaton sector fields (essentially droplets of Bose condensate held together by an attractive interaction) was observed. The possibility of the formation of 
 such non-topological solitons  at the end of SUSY hybrid inflation was first suggested in \cite{icf} (where they were called 'inflation condensate lumps') and their formation was also observed in a numerical simulation of the growth of perturbations of the symmetry-breaking field in SUSY hybrid inflation models 
\cite{cpr}. Similar oscillon states have been observed in numerical simulations     
 of the growth of scalar field fluctuations along SUSY flat directions \cite{kk} and in a chaotic inflation model \cite{ke1} based on the 
SUSY flat direction scalar potential \cite{qbo} . In those cases the oscillons were observed to subsequently decay to Q-ball anti-Q-ball pairs. This is a natural possibility since the Q-ball is an energetically preferred stable state, to which the unstable neutral oscillons will naturally evolve via the growth of perturbations of the phase of the complex scalar field. A similar decay of hybrid inflaton oscillons to Q-balls might therefore be expected if stable Q-balls made of inflaton sector fields exist, resulting in a Q-ball dominated post-inflation era. This would have important consequences for 
reheating \cite{icf,mb1} and post-inflation scalar field dynamics \cite{fdd}. 
           
     In a previous paper we demonstrated the existence of Q-ball solutions of the field equations of the D-term 
inflation model \cite{mq1}. It was noted that, for a particular choice of couplings, the 
Q-ball solutions in the D-term inflation model are equivalent to Q-balls in the 
minimal F-term inflation model, based on a single symmetry breaking field which carries no gauge charge. 
On the other hand, the most commonly studied F-term inflation model is that based on a pair 
of symmetry-breaking fields,  $\Phi$ and $\overline{\Phi}$, transforming as a conjugate pair under a gauge group. 
(In the following we will refer to this as standard F-term inflation.) The Q-ball solutions in this model are not equivalent to those in D-term inflation. 

           In this paper we present a detailed analysis of Q-ball solutions in the standard F-term inflation model.  
We will show that the Q-ball solutions for any superpotential coupling, $\kappa$, can be made equivalent to those with $\kappa = 1$ 
by rescaling the space coordinates. As a result, we are able to study the properties of F-term inflation Q-balls in a completely general way.  We will focus on the case of global SUSY throughout. The effect of supergravity (SUGRA) corrections in realistic models will be briefly discussed in our conclusions.

      The paper is organised as follows. In Section 2 we derive the equations which minimize the energy for a fixed global charge ('Q-ball equations') for the standard  F-term inflation model. In Section 3 we show that the Q-ball solutions for any 
superpotential coupling can be obtained by rescaling the space coordinates and solving the field equations for the case $\kappa = 1$. 
In Section 4 we present numerical solutions of the standard F-term inflation Q-ball equations and give the full range of 
solutions for the case $\kappa = 1$.  In Section 5 we present our conclusions and discuss the possible role of F-term inflation Q-balls in cosmology.

\section{Q-Ball Equations of the Standard F-term Inflation Model}

           The most commonly studied form of F-term inflation model has the superpotential \cite{lr,fti}
\be{n1}  W = \kappa S \left(\overline{\Phi} \Phi - \mu^{2}\right)    ~,\ee
where $S$ is the inflaton and $\Phi$ and $\overline{\Phi}$ are the symmetry-breaking fields.
 We may choose $\mu^{2}$ to be real and positive. The symmetry-breaking fields typically represent 
oppositely charged gauge multiplets in realistic models. The scalar potential is then  
\be{n2} V = \kappa^{2} \left| \overline{\Phi} \Phi - \mu^{2} \right|^{2} + \kappa^{2} |S|^{2} \left( 
\left| \overline{\Phi} \right|^{2} + |\Phi|^{2} \right)  + \frac{g^{2}}{2} \left( \left| \overline{\Phi} \right|^{2} 
- \left| \Phi \right|^{2} \right)^{2}        ~,\ee
where $\Phi$ and $\overline{\Phi}$ now represent the components of the gauge multiplets which gain an expectation value. 
In this we have included a generic D-term for the case where $\Phi$, $\overline{\Phi}$ transform under a gauge group. 
The vacuum state will correspond to $S = 0$ and $\overline{\Phi} \Phi  = \mu^{2}$. 
In the case where there is a D-term, the vacuum will be such that 
$ \left| \overline{\Phi} \right| = \left| \Phi \right| $. We can choose $\Phi$ to be real and positive, in which case 
the vacuum expectation values (VEVs) will correspond to $ \overline{\Phi}  =  \Phi  = \mu $.  
We will focus on this case in the following.

     A superpotential proportional to $S$, together with a minimal 
K\"ahler potential term, is essential in order to avoid the $\eta$-problem in F-term
inflation models once SUGRA corrections are included \cite{eta}. 
A superpotential which is proportional to $S$ implies that there is a 
global U(1) R-symmetry, under which only the inflaton transforms, and an associated global charge. 
The existence of a conserved inflaton charge, $Q_{S}$, implies the possibility of Q-ball solutions of the 
F-term inflation field equations.

For a model with a single complex scalar field, $\Psi$,
and a global U(1) symmetry under which  $\Psi \rightarrow e^{i \alpha} \Psi$,  the Q-ball configuration is
derived by minimizing the functional
\be{n7} E_{\omega}(\dot{\Psi},\Psi,\omega) = E + \omega \left(Q - \int d^{3}x \rho_{Q}\right)   ~,\ee
where $\omega$ is a Lagrange multiplier fixing the charge,  
$E$ is the total energy of the field configuration
\be{n8} E = \int \left[ |\dot{\Psi}|^{2} + |\underline{\nabla} \Psi|^{2}
+ V(|\Psi|) \right] \; d^{3}x  ~\ee 
and $\rho_{Q}$ is the global charge density
\be{n9} \rho_{Q} = i( \dot{\Psi}^{\dagger}\Psi - \Psi^{\dagger} \dot{\Psi})  ~.\ee
This minimizes the energy for a fixed global charge, $Q$. $E_{\omega}$ may be equivalently written as 
\be{n10} E_{\omega}(\dot{\Psi},\Psi,\omega)  = \int \left[ |\dot{\Psi}
 - i \omega \Psi|^{2} + 
 |\underline{\nabla}\Psi|^{2} + V(\Psi) - \omega^{2} |\Psi|^{2} \right] \; d^{3}x  + \omega Q  
 ~.\ee
Minimizing with respect to $\dot{\Psi}$ implies that $\Psi(\vec{x},t)
 = \Psi(\vec{x}) e^{i \omega t}$. Substituting this into \eq{n10} gives 
\be{n11} E_{\omega}(\Psi(\vec{x}),\omega) = \int \left[
 |\underline{\nabla}\Psi(\vec{x})|^{2}
 + V(\Psi(\vec{x})) - \omega^{2} |\Psi(\vec{x})|^{2} \right] \; d^{3}x  + \omega Q  ~.\ee
Extremizing this with respect to $\Psi(\vec{x})$ then implies that
\be{n12}  \underline{\nabla}^{2} \Psi(\vec{x}) = \frac{\partial
 V_{\omega}\left(\Psi(\vec{x})\right)}{\partial \Psi^{\dagger}}   ~\ee
where $V_{\omega} = V(\Psi) - \omega^{2} |\Psi|^{2}$.   
$\Psi(\vec{x})$ could still have a space-dependent complex phase, 
$\theta(\vec{x})$. 
However, if $V(\Psi) = V(|\Psi|)$ (as it must when $\Psi$ transforms under a $U(1)$ symmetry), 
then \eq{n12} is generally
 minimized by the choice $\theta = $ {\it constant}, 
which may be chosen such that $\Psi(\vec{x})$ is real.   
A minimum energy configuration should be spherically symmetric. 
Therefore, with $\Psi(\vec{x}) = \psi(r)/\sqrt{2}$  (where $\psi(r)$ is real), \eq{n12} becomes 
\be{n13}  \frac{ \partial^{2} \psi}{\partial r^{2}}  +\frac{2}{r} \frac{\partial
 \psi}{\partial r} = \frac{ \partial V}{\partial \psi} - \omega^{2} \psi    ~.\ee
We will refer to this as the Q-ball equation. The solutions of \eq{n13} should satisfy the boundary 
conditions that the field tends to the vacuum as 
$r \rightarrow \infty$ and that $\partial \psi/\partial 
r \rightarrow 0$ as $r \rightarrow 0$. 
On physical grounds we expect that there will be a unique Q-ball solution of this form
for a given global charge.  The resulting solution will be a stable Q-ball if $E/Q < m_{\Psi}$, 
since it would then require additional energy to remove the conserved global charge into the vacuum. 

       The procedure for solving \eq{n13} numerically to obtain Q-ball solutions for a given value of 
$\omega$ is as follows. We first input a value for $\psi(r=0) \equiv \psi_{o}$, impose 
the boundary condition that $\partial \psi/\partial r = 0$ at $r = 0$, 
and solve the Q-ball equation for $\psi(r)$. 
We then vary $\psi_{o}$ until we obtain a solution which asymptotically approaches the 
vacuum as $r \rightarrow \infty$. This will correspond to a possible Q-ball solution. 
Finally, we calculate $E$ and $Q$ for this solution and check that $E/Q < m_{\Psi}$.  
In practice we can only check the asymptotic approach to the vacuum out to a finite radius, $r_{end}$. 
So long as $r_{end}$ is large compared with the dynamical scale of the scalar field, $m_{\Psi}^{-1}$, 
this will give a good approximation to the true Q-ball solution. Larger values of $r_{end}$ require 
progressively more accurate values of $\psi_{o}$. 
     
         The above analysis generalises to the case of SUSY 
hybrid inflation with two or more complex scalar fields. 
The Q-ball equations generally become a system of coupled
equations for the inflaton, $s$, and
symmetry-breaking field, $\phi$, with a single Lagrange
multiplier associated with the inflaton. 
The numerical procedure is as before, except that we must scan through 
values of $s_{o}$ and $\phi_{o}$, looking for solutions which satisfy the requirement that 
$s$ and $\phi$ asymptotically approach their VEV as $r \rightarrow \infty$.

         For the case of standard F-term inflation there are three complex scalar fields. 
However, we can only associate a Lagrange multiplier with fields which have 
vacuum expectation value equal to zero. This follows since the energy minimization 
with a Lagrange mulitplier $\gamma$ for a field $\Phi$  
which transforms under a global U(1) symmetry implies that the Q-ball solution must be of the form 
$\Phi(\vec{x}, t) =  \phi(r)e^{i \gamma t}/\sqrt{2}$. This is only compatible with a
constant VEV as $r \rightarrow \infty$ if $\phi(r) \rightarrow 0$.          
Therefore in the case of F-term inflation, where $\Phi$ and $\overline{\Phi}$ have non-zero VEVs,
 we can only associate a Lagrange multiplier with the $S$ field.

   Thus we will look for Q-ball solutions with a global charge $Q_{S}$ associated with 
the R-symmetry under which $S \rightarrow e^{i \alpha} S$. The energy functional is then 
\be{n15} E_{\omega} = \int  \left[ |\dot{S} - i \omega S|^{2} + |\underline{\nabla} S|^{2} +
|\dot{\Phi}|^{2} + |\underline{\nabla} \Phi|^{2} +
|\dot{\overline{\Phi}}|^{2} + |\underline{\nabla} \overline{\Phi}|^{2}
+ V_{\omega}(|S|, \Phi, \overline{\Phi}) \right] \; d^{3}x   + \omega Q_{S}   ~,\ee 
where
\be{n18} 
V_{\omega}(|S|, \Phi, \overline{\Phi}) = V(|S|, \Phi, \overline{\Phi}) - \omega^{2} |S|^{2}  ~,\ee
with $V(|S|, \Phi, \overline{\Phi})$ given by \eq{n2}. 
Minimizing $E_{\omega}$ with respect to the time derivatives then implies that 
$S(\vec{x},t)= S(\vec{x})e^{i \omega t}$, $\Phi(\vec{x},t) =
\Phi(\vec{x})$ and $\overline{\Phi}(\vec{x},t) = \overline{\Phi}(\vec{x})$. 
\eq{n15} then becomes
\be{n17} E_{\omega} = \int \left[ |\underline{\nabla} S(\vec{x})|^{2} +
|\underline{\nabla} \Phi (\vec{x})|^{2} + |\underline{\nabla} \overline{\Phi}(\vec{x})|^{2}
+ V_{\omega}(|S|, \Phi, \overline{\Phi}) \right] \; d^{3}x + \omega Q_{S}   ~.\ee
The potential is explicitly dependent upon the phase of $\Phi$ and $\overline{\Phi}$. However, by inspection of \eq{n2}, 
the minimum of $E_{\omega}$ with $\mu^{2}$ real and positive will generally correspond to $\Phi$
and $\overline{\Phi}$ real and positive. Therefore with $S(\vec{x}) = s(r)/\sqrt{2}$, 
$\Phi(\vec{x}) = \phi(r)/\sqrt{2}$, $\overline{\Phi}(\vec{x}) = \overline{\phi}(r)/\sqrt{2}$, the 
Q-ball equations analogous to \eq{n13} from extremizing \eq{n17} are  
\be{n19}  \frac{ \partial^{2} s}{\partial r^{2}}  + \frac{2}{r} \frac{\partial s}{\partial
 r} = \frac{\kappa^{2}}{2}
 \left(\phi^{2} + \overline{\phi}^{2}\right) s - \omega^{2} s    ~,\ee
\be{n20}  \frac{ \partial^{2} \phi}{\partial r^{2}}  +\frac{2}{r} \frac{\partial
 \phi}{\partial r} =  
\kappa^{2} \left( \frac{\overline{\phi} \phi}{2} - \mu^{2} \right) \overline{\phi} + 
\frac{1}{2} \kappa^{2} s^{2} \phi + \frac{g^{2}}{2} \left(\phi^{2}
 - \overline{\phi}^{2} \right) \phi    ~\ee 
and 
\be{n21}  \frac{ \partial^{2} \overline{\phi}}{\partial r^{2}}  +\frac{2}{r} \frac{\partial
 \overline{\phi}}{\partial r} =  
\kappa^{2} \left( \frac{\overline{\phi} \phi}{2} - \mu^{2} \right) \phi + 
\frac{1}{2} \kappa^{2} s^{2} \overline{\phi} + \frac{g^{2}}{2} \left(\overline{\phi}^{2}
 - \phi^{2} \right) \overline{\phi}    ~.\ee  
Since in the case with a D-term the VEV of $\Phi$ and $\overline{\Phi}$ are both equal to $\mu$, 
we will look for Q-ball solutions with $\phi(r) = \overline{\phi}(r)$. In this case the equations reduce to 
\be{n22}  \frac{ \partial^{2} s}{\partial r^{2}}  + \frac{2}{r} \frac{\partial s}{\partial
 r} =  \kappa^{2} \phi^{2} s - \omega^{2} s    ~\ee
and
\be{n23}  \frac{ \partial^{2} \phi}{\partial r^{2}}  +\frac{2}{r} \frac{\partial
 \phi}{\partial r} =  
\kappa^{2} \left( \frac{\phi^{2}}{2} - \mu^{2} \right)\phi + 
\frac{1}{2} \kappa^{2} s^{2} \phi ~.\ee  
The energy and charge of the resulting Q-ball soultion are given by
\be{n23a}  E = \int 4 \pi r^{2}  \left[ \frac{1}{2} \left( \frac{\partial s}{\partial
 r}\right)^{2} +  \left( \frac{\partial \phi}{\partial r}\right)^{2} 
+ \frac{\omega^{2} s^{2}}{2} + V(s,\phi) \right] dr
~,\ee
where 
\be{n23b} V(s, \phi) = \kappa^{2} \left( \frac{\phi^{2}}{2}
 - \mu^{2} \right)^{2} + \frac{\kappa^{2}}{2} s^{2} \phi^{2}   ~,\ee
and 
\be{n23c} Q_{S} \equiv \int \rho_{Q_{S}} d^{3}x = \omega \int 4 \pi r^{2} s^{2}dr    ~,\ee
where $ \rho_{Q_{S}} = i \left(\dot{S}^{\dagger}S - S^{\dagger} \dot{S} \right)$. 
There is no factor of 1/2 in front of $(\partial \phi/\partial r)^{2}$
contribution to $E$, since this includes the energy associated with $\overline{\phi}$ in the case where 
$\overline{\phi} = \phi$.

\section{Scaling Property of the Q-Ball Equations}

         The two free parameters of the standard F-term inflation superpotential are 
the mass scale, $\mu$, and the superpotential coupling, $\kappa$. The mass scale may be
eliminated by using units such that $\mu = 1$. We can avoid having to produce a set of Q-ball solutions
for each value of $\kappa$ by rewriting the Q-ball equations in terms of 
$\tilde{r} = \kappa r$. The Q-ball equatons then become     
\be{n24}  \frac{ \partial^{2} s}{\partial \tilde{r}^{2}}  + \frac{2}{\tilde{r}} \frac{\partial s}{\partial
 \tilde{r}} =  \phi^{2} s - \tilde{\omega}^{2} s    ~\ee
and
\be{n25}  \frac{ \partial^{2} \phi}{\partial \tilde{r}^{2}}  +\frac{2}{\tilde{r}} \frac{\partial
 \phi}{\partial \tilde{r}} =  
\left( \frac{\phi^{2}}{2} - 1 \right)\phi + 
\frac{1}{2} s^{2} \phi ~,\ee
where $\tilde{\omega} = \omega/\kappa$. These equations have the same form as the Q-ball equations for the case 
$\kappa = 1$ in $\mu = 1$ units.   
The energy and charge are given by
\be{n26}  E = \frac{1}{\kappa} \int 4 \pi \tilde{r}^{2} \left[ \frac{1}{2} \left( \frac{\partial s}{\partial
 \tilde{r}}\right)^{2} +  \left( \frac{\partial \phi}{\partial \tilde{r}}\right)^{2} 
+ \frac{\tilde{\omega}^{2} s^{2}}{2} + \tilde{V}(s,\phi) \right]  d\tilde{r}  \equiv \frac{\tilde{E}}{\kappa} 
~,\ee
where 
\be{n27} \tilde{V}(s, \phi) = \left( \frac{\phi^{2}}{2}
 - 1 \right)^{2} + \frac{1}{2} s^{2} \phi^{2}   ~,\ee
and 
\be{n28} Q_{S} = \frac{\tilde{\omega}}{\kappa^{2}}  \int 4 \pi \tilde{r}^{2} s^{2} d\tilde{r} \equiv  \frac{\tilde{Q}_{S}}{\kappa^{2}}    ~.\ee
$\tilde{E}$ and $\tilde{Q}_{S}$ are the energy and charge of the corresponding $\kappa = 1$ Q-ball. The inflaton mass is given by $m_{S} = \sqrt{2} \kappa \mu$. Therefore the Q-ball stability condition, $E/Q_{S}m_{S} < 1$, is independent of $\kappa$.

         Since all the Q-balls for an arbitrary value of $\kappa$ have a one-to-one mapping to those
 with $\kappa = 1$, the full set of Q-ball solutions for the case $\kappa = 1$ will constitute a complete solution
 for the range of possible Q-balls in standard F-term inflation.

\section{Numerical Q-ball Solutions}

 In Table 1 we give the set of $\kappa = 1$ Q-ball solutions we have generated in order to analyze the general properties of F-term inflation Q-balls. 
We use $\mu = 1$ units throughout. The inflaton mass is then $m_{S} = \sqrt{2} \kappa$ and the VEV of $\phi$ equals $\sqrt{2}$. The solutions were 
generated using a second-order Runge-Kutta routine with radial increment $\Delta r = 0.01$.

 \begin{table}[h]
 \begin{center}
 \begin{tabular}{|c|c|c|c|c|c|c|c|}
	\hline  \bf{$\omega$} & $s_{o}$ & $\phi_{o}$ &  $E$ & $Q_{S}$ & $E/Q_{S}m_{S}$ & $R$ & $r_{end}$ \\ 
	\hline  $0.58$ & $7.7891$ & $1.03 \times 10^{-6}$ & $2756.42$ & 3654.07& $0.533$ & $5.06$ & 6.40\\
	\hline  $0.6$ & $7.2700$ & $5.40 \times 10^{-6}$ & $2370.57$ & 3006.14& $0.557$ & $4.92$ & 8.0\\
	\hline  $0.65$ & $6.6699$ & $5.24 \times 10^{-5}$ & $1855.8$ & 2180.4& $0.602$ & $4.57$ & 8.0\\
	\hline  $0.7$ & $6.1518$ & $3.16 \times 10^{-4}$ & $1479.41$ & 1621.36 & $0.645$ & $4.28$ & 8.0\\
	\hline  $0.8$ & $5.2854$ & $4.1436 \times 10^{-3}$ & $980.86$ & 951.66 & $0.729$ & $3.83$ & 8.0\\
	\hline  $0.9$ & $4.57364$ & $2.3758 \times 10^{-2}$ & $681.07$ & 596.90 & $0.807$ & $3.51$ & 8.0\\
	\hline  $1.0$ & $3.95008$ & $0.08160$ & $491.15$ & 396.03& $0.877$ & $3.32$ & 8.0\\
	\hline  $1.1$ & $3.35932$ & $0.202129$ & $366.43$ & 276.78 & $0.936$ & $3.24$ & 8.0\\
	\hline  $1.17$ & $2.93500$ & $0.334867$ & $304.56$ & 222.17& $0.969$ & $3.28$ & 8.0\\
	\hline  $1.25$ & $2.38244$ & $0.552551$ & $255.58$ & 181.67& $0.995$ & $3.61$ & 6.90\\
      \hline  $1.30$ & $2.0631$ & $0.69888$ & $222.45$ & 156.02& $1.013$ & $3.60$ & 6.30\\
	\hline     
 \end{tabular}
 \caption{\footnotesize{Properties of F-term inflation Q-ball solutions with $\kappa =1$.}}  
 \end{center}
 \end{table}

                              In Figure 1 we show $s(r)$ and $\phi(r)$ for a typical $\kappa = 1$ F-term
inflation Q-ball. In this case ($\omega = 1.1$) we find that $s(r)$ can be approximately fitted by a Gaussian, 
$s(r) = s_{o} \exp(-r^{2}/\hat{R}^{2})$ with $s_{o} = 3.36$ and $\hat{R} = 2.24$. This is a good fit for $r \lae 4.0$, but for larger $r$ the
Gaussian gives a value of $s(r)$ somewhat smaller than the numerical solution (by a factor of 10 at $r = 6.0$). Nevertheless, the Gaussian provides a useful analytical expression for $s(r)$, which can be used when studying the cosmology of Q-balls \cite{fdd}.

  \begin{figure}[h] 
                    \centering                   
                    \includegraphics[width=0.50\textwidth, angle=-90]{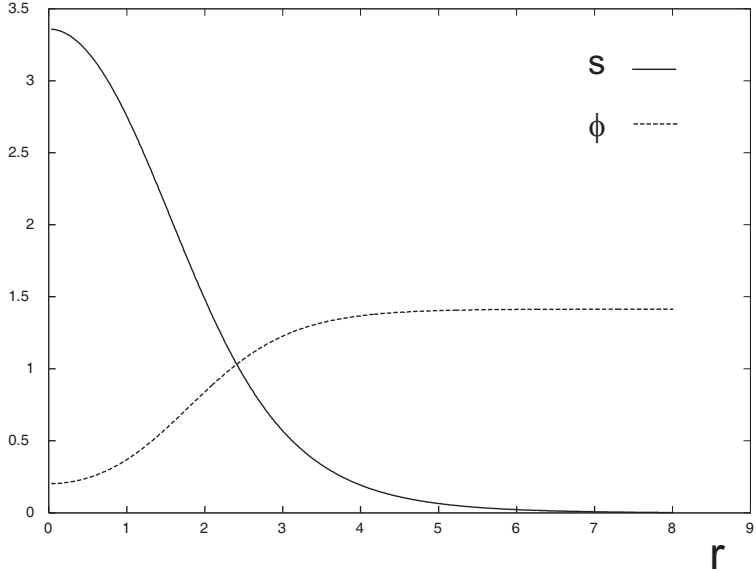}
                    \caption{\footnotesize{F-term Inflation Q-ball for the case $\kappa = 1$, $\omega = 1.1$.}}
                    \end{figure}

  \begin{figure}[hp] 
                     \centering                                     
                     \includegraphics[width=0.50\textwidth, angle=-90]{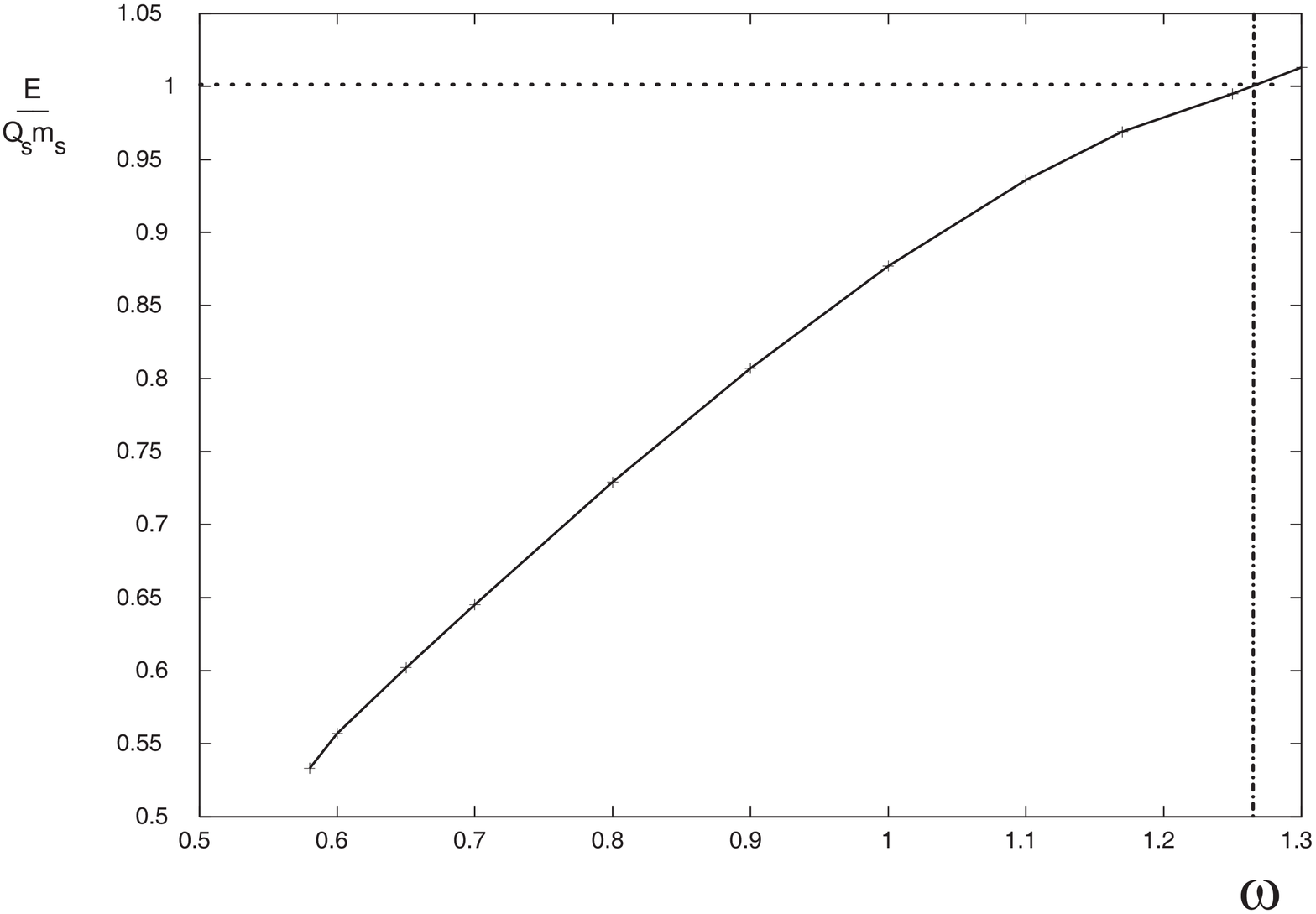}
                     \caption{\footnotesize{$\frac{E}{Q_{S} m_{S}}$ versus $\omega$ for $\kappa = 1$ Q-balls. The Q-ball stability condition,
                     $\frac{E}{Q_{S} m_{S}} < 1$, implies that $\omega \lae 1.26$.}}
                     \includegraphics[width=0.50\textwidth, angle=-90]{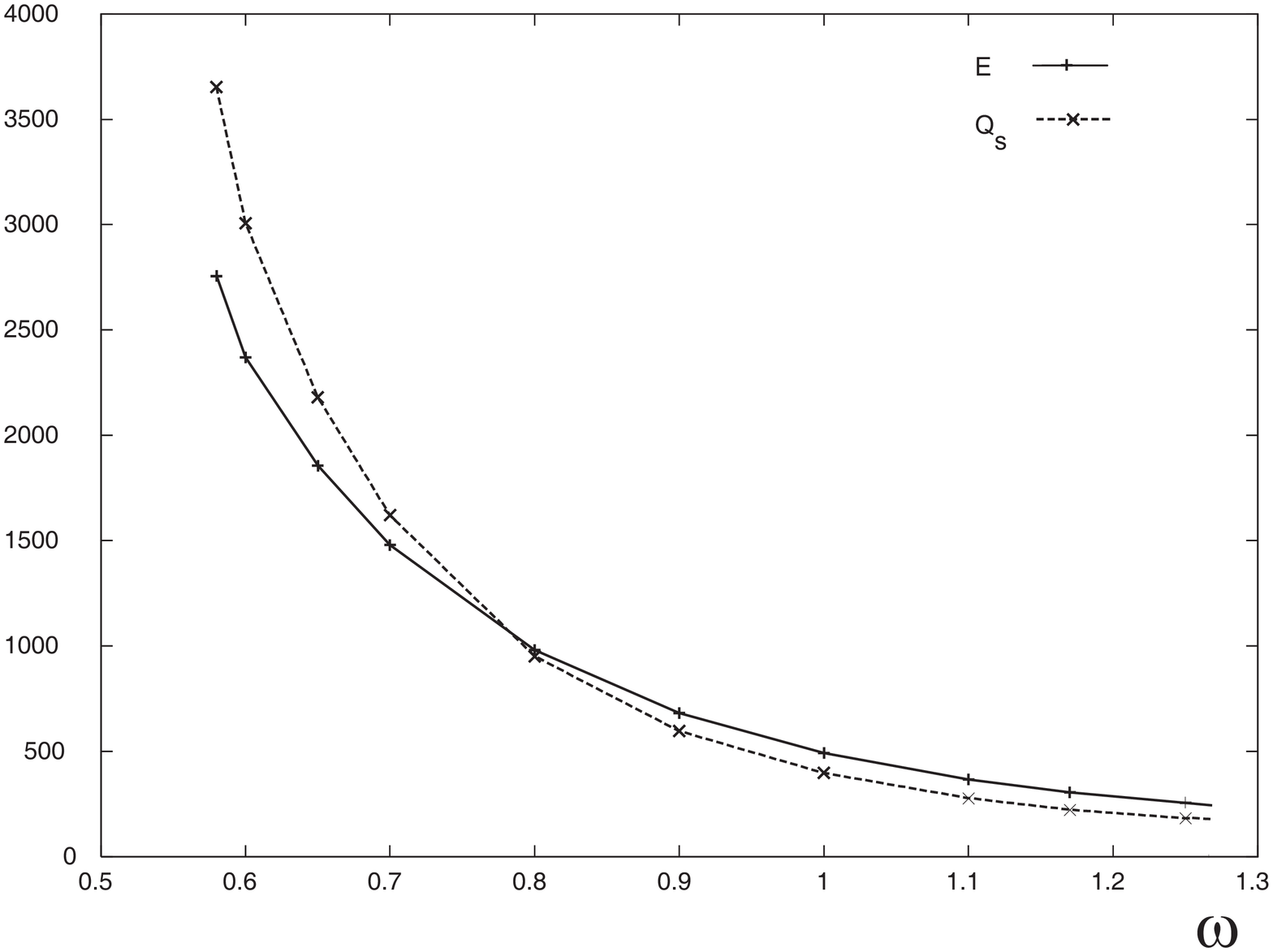}
                     \caption{\footnotesize{Energy $E$ and charge $Q_{S}$ versus $\omega$ for $\kappa = 1$ Q-balls. Both diverge as $\omega \rightarrow 
                     0.57$.}}
                     \end{figure}

 \begin{figure}[hp] 
                    \centering                   
                    \includegraphics[width=0.50\textwidth, angle=-90]{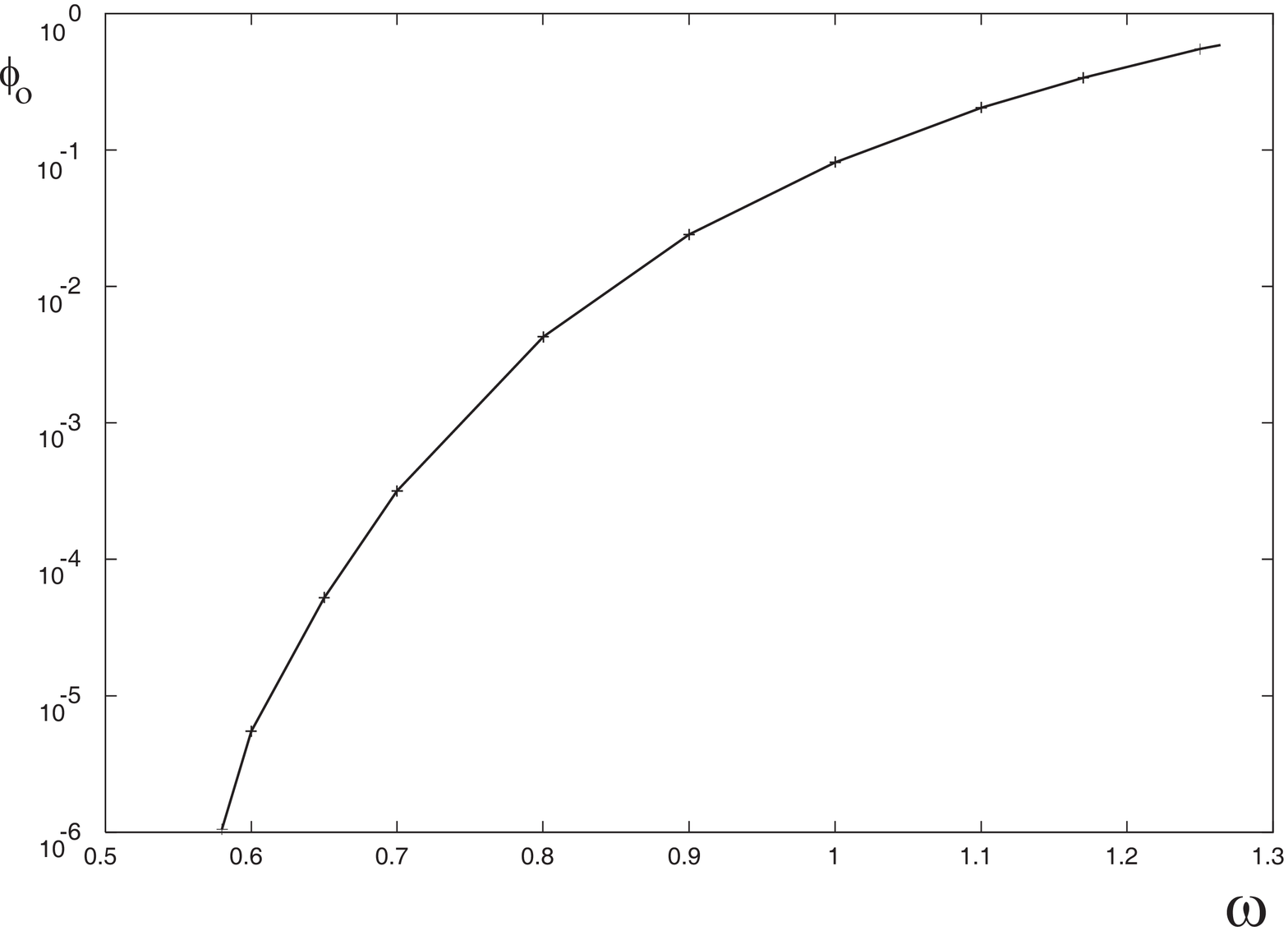}
                    \caption{\footnotesize{$\phi_{o}$ versus $\omega$ for $\kappa = 1$ Q-balls}}
                    \includegraphics[width=0.50\textwidth, angle=-90]{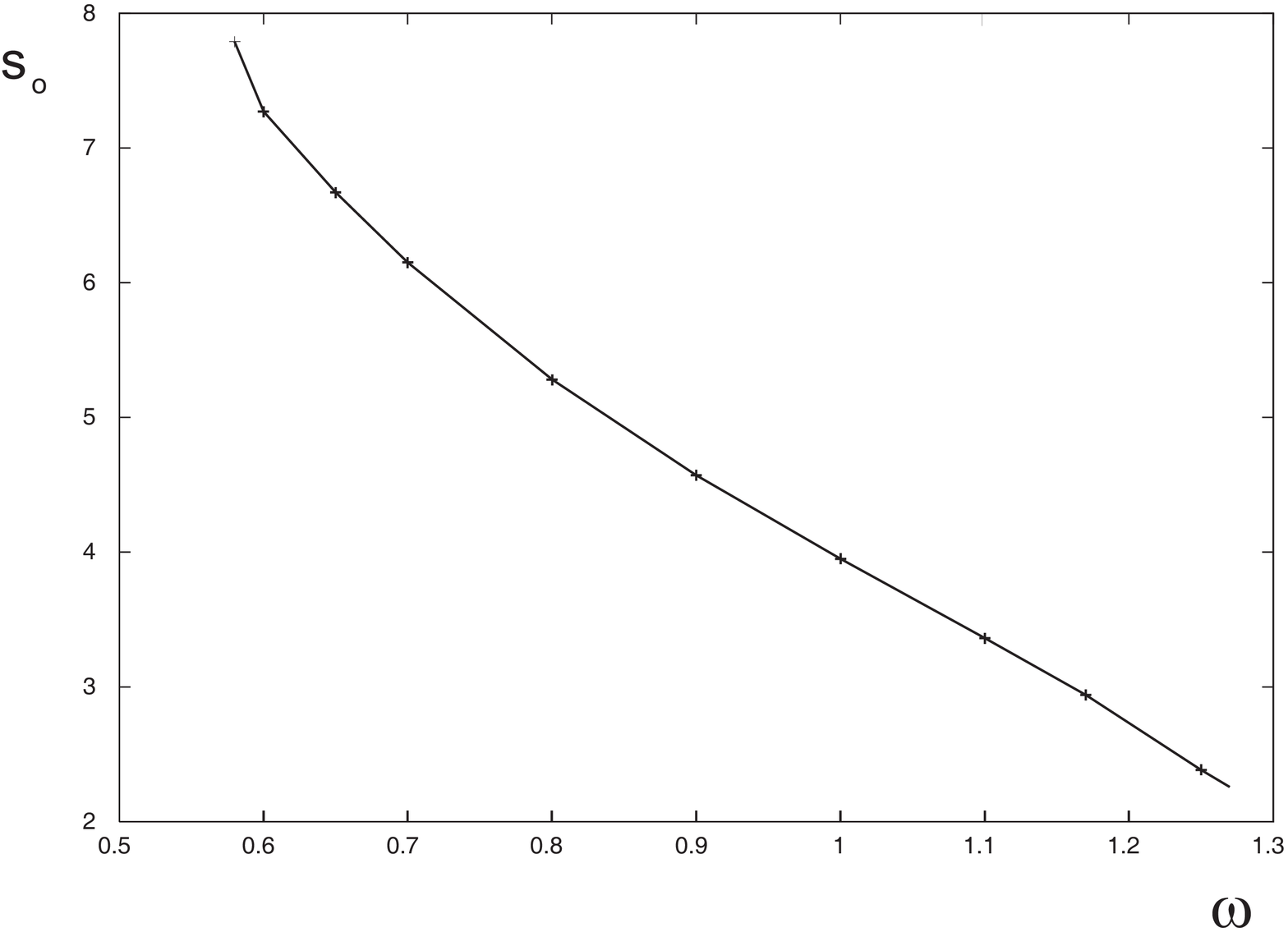}
                    \caption{\footnotesize{$s_{o}$ versus $\omega$ for $\kappa = 1$ Q-balls}}
                    \end{figure}

 \begin{figure}[h] 
                    \centering                   
                    \includegraphics[width=0.50\textwidth, angle=-90]{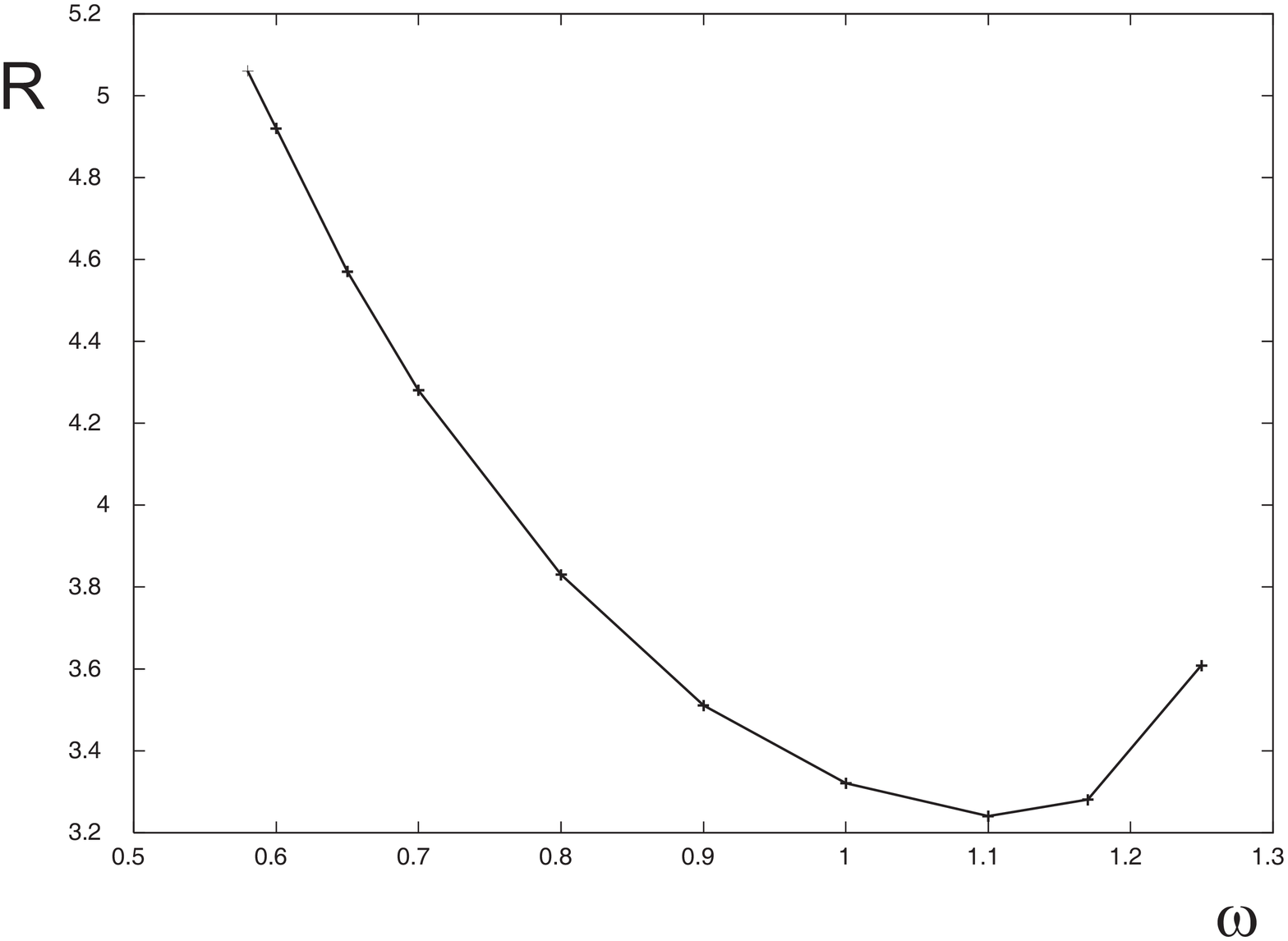}
                    \caption{\footnotesize{Radius versus $\omega$ for $\kappa = 1$ Q-balls}}
                    \end{figure}

               In Figure 2 we show the ratio $E/Q_{S} m_{S}$ as a function of $\omega$. As discussed below, there is a lower limit on $\omega$, approximately 0.57, at which value the energy and charge of the Q-ball diverges. An upper limit on $\omega$ is imposed by the requirement that 
$E/Q_{S} m_{S} < 1$ for a stable Q-ball. From Figure 2 we see that this requires that $\omega \lae 1.26$. (We will apply this upper limit to the 
remaining figures.)

   In Figure 3 we show the energy and charge of the Q-ball solutions as a function of $\omega$. Both $E$ and $Q_{S}$ display vertical asymptotic behaviour as $\omega$ approaches $0.57$, therefore Q-balls of arbitrarily large energy and charge are possible. The upper bound on $\omega$ coming from Q-ball stability imposes a lower bound on the energy and charge for a stable Q-ball solution to exist, $E_{min} \approx 250$ and $Q_{min} \approx 180$.

      In Figure 4 we show $\phi_{o}$ as a function of $\omega$. We find that, as $\omega$
decreases, the value of $\phi_{o}$ becomes small and asymptotically approaches zero as $\omega \rightarrow 0.57$. 
The extremely small and precise values of $\phi_{o}$ required as $\omega$ approaches 0.57 makes it difficult
to find Q-ball solutions in this limit. The value of $s_{o}$ also begins to increase rapidly as $\omega \rightarrow 0.57$, as shown in Figure 5.

         In Figure 6 we show the Q-ball radius, $R$, as a function of $\omega$, where we have defined $R$ to be the radius within which 90$\%$ of its energy resides. We find that the radius of the Q-balls is relatively
insensitive to the energy and charge, varying from 3.27 to 5.06 over the entire range of Q-ball solutions.

              For the case of Q-balls with $\kappa < 1$, the effect of rescaling the
$\kappa = 1$ coordinates will be to increase the lower bound on the energy to $E_{min} \approx 250/\kappa$ 
and the charge to $Q_{min} \approx 180/\kappa^{2}$. The corresponding Q-balls will have
a larger physical radius, $R = R(\kappa = 1)/\kappa$. The range of $\omega$ for which stable Q-ball solutions exist becomes $0.57 \kappa \lae \omega \lae 1.26 \kappa$.

\section{Conclusions and Discussion}

                   In this paper we have studied the complete range of possible Q-ball solutions in the standard F-term inflation model. 
The solutions are parameterized by the frequency of rotation of the inflaton in the complex plane, $\omega$. It is useful to restate our main results in terms of the physical inflaton mass ($m_{S} = \sqrt{2} \kappa$ in $\mu = 1$ units). For a superpotential coupling $\kappa$, stable Q-balls exist for $0.40  m_{S} \lae \omega \lae 0.89 m_{S}$. As $\omega \rightarrow 0.40 m_{S}$, 
the charge and energy of the Q-ball diverge. There is a lower bound on the charge and energy for which a stable Q-ball exists, 
$Q_{min} \gae 180/\kappa^{2}$ and $E_{min} \gae 180 m_{S}/\kappa^{2}$. The radius of the Q-ball, within which 90$\%$ of its energy resides, is between $4.62 m_{S}^{-1}$ and $7.16 m_{S}^{-1}$.

           The existence of Q-ball solutions of the standard F-term inflation model, combined with the rapid growth of quantum scalar field fluctuations 
at the end of inflation, implies that Q-ball formation at the end of F-term inflation is a possibility. 
What can we deduce about F-term inflation Q-ball cosmology, should 
their formation at the end of inflation be confirmed in future 
numerical simulations? The typical energy and charge of such Q-balls, assuming 
that $\omega$ is not close to its upper or lower limit, would be $E \approx 10^{3} m_{S}/\kappa^{2}$ and $Q_{S} \approx 10^{3}/\kappa^{2}$. It was recently shown that cosmic microwave background constraints impose limits on the range of $\kappa$ in F-term inflation models,  
$10^{-7} \lae \kappa \lae 10^{-2}$ \cite{jean}. Within this range, the global charge carried by the Q-balls can be very large, up to 
$Q_{S} \approx 10^{17}$. In the case where gravity-mediated SUSY breaking A-terms exist during inflation, the lower bound becomes $\kappa \gae 10^{-5}$ \cite{ss}, corresponding to Q-ball charges up to $Q_{S} \approx 10^{13}$. Such large charge Q-balls would be expected to have interesting consequences for reheating, delaying the process relative to the 
case of a conventional perturbatively decaying homogeneous inflaton condensate \cite{qbo,enqreh}. 
The consequences of inhomogeneous reheating occuring
via Q-ball decay would be particularly interesting if it occurs close to 
a physically significant temperature, such as that of the electroweak 
phase transition or dark matter freeze-out.

      The existence of a lower bound on the charge of a stable Q-ball has interesting
 implications for the Q-ball decay process. The inflatons making up the Q-ball would decay
 perturbatively until the charge drops below $Q_{min}$, at which point the Q-ball would rapidly expand
 due to its gradient energy and dissociate into inflatons.

               It should be emphasized that the existence of Q-ball solutions 
is not dependent upon the existence of a symmetry-breaking phase transition     
in hybrid inflation models. For example, it is straightforward to show that Q-ball solutions also exist
in smooth hybrid inflation models \cite{shi}, where inflation ends without any phase transition \cite{shiq}.

         In this paper we have considered F-term inflation Q-balls in the limit of global SUSY. 
The F-term inflation Q-ball solutions are stable as a result of a global U(1) symmetry, which must correspond
to an R-symmetry in order to be consistent with the $\kappa \mu^{2} S$ term in the superpotential. 
However, in realistic models we must also consider the effect of SUGRA corrections. 
If SUGRA with hidden sector R-symmetry breaking is considered then
there will be U(1)-breaking SUGRA corrections which could destabilise the Q-balls\footnote{We 
note that in the case of D-term inflation the 
Q-balls \cite{mq1} are stable due to a global U(1) symmetry of the superpotential and so are
not affected by R-symmetry breaking SUGRA corrrections.}.
For example, in the case of gravity-mediated SUSY breaking, 
an R-symmetry breaking A-term of the form $V_{A} \approx m_{3/2} \kappa \mu^{2} S + h.c.$
will generally be added to the scalar potential \cite{ss}, which would add a constant term to the 
right hand side of the $S$ field equation (\eq{n22}) of the form $m_{3/2} \kappa \mu^{2}$. 
The magnitude of $s$ and $\phi$ inside the Q-ball is typically of the order of $\mu$, so the ratio of
the U(1)-breaking A-term to the U(1)-symmetric terms in \eq{n22} is approximately $m_{3/2}/m_{S}$.
Since the inflaton mass, $m_{S}$, is typically many orders of magnitude larger than $m_{3/2}$, 
the A-term will make a very small contribution to the $S$ field equation compared with the U(1)-symmetric terms. 
In this case there will be a time-dependent oscillon solution with properties very close to those of a Q-ball.
Given the small size of the A-term and the inherent stability of oscillon solutions \cite{osc}, we would expect the lifetime of the Q-ball-like objects to be very long compared with the dynamical timescale $m_{S}^{-1}$.  
However, a numerical solution for the effect of the A-term on the Q-ball lifetime will be necessary in order
to find out whether it could dominate over perturbative inflaton decay.

             Beyond cosmology, the F-term inflation Q-ball provides an interesting example
of a non-topological soliton in the context of a physically-motivated scalar field theory. It
would be interesting to try to understand analytically some of the features which we have found
numerically, such as the lower bound on the charge and energy and the finite range of $\omega$. 
The properties of the Q-balls might also shed light on the dynamics of the related oscillons which 
occur in the case of a real inflaton field. For example, the existence of a lower bound on the Q-ball 
energy to form a stable Q-ball might imply a similar lower bound on the oscillon energy. The advantage 
of the Q-ball is that it is possible solve exactly for its time-dependence, whereas oscillons generally 
have a complicated time-dependence which is difficult to study numerically \cite{osc}.

\end{document}